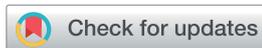

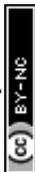

**PAPER**

Check for updates

## Geometry-induced enhancement factor improvement in covered-gold-nanorod-dimer antennas

Iván A. Ramos,[a] L. M. León Hilario, [ID] *[a] María L. Pedano[b] and Andres A. Reynoso [ID] *[b]

Illuminated gapped-gold-nanorod dimers hold surface plasmon polaritons (SPPs) that can be engineered, by an appropriate choice of geometrical parameters, to enhance the electromagnetic field at the gap, allowing applications in molecular detection *via* surface-enhanced Raman spectroscopy (SERS). Envisioning hybrid devices in which the SERS spectroscopy of molecules in the gap is complemented by electrical measurements, it arises the question of designing efficient geometries to contact the nanorods without decreasing the enhancement factor (EF) of the nanoantenna, *i.e.*, the figure of merit for SERS spectroscopy. Within this framework we theoretically study the feasibility to fabricate designs based on covering with gold the far-from-the-gap areas of the dimer. We show that by tuning the geometrical parameters of the designs these systems can reach enhancement factors larger than the best achieved in the uncovered dimer: this supremacy survives even in the presence of dimer asymmetries and vacancies at the interfaces between the nanorods and the covering layers. Our results show that geometrical modifications away from the gap can improve the optical response at the gap, thus enabling the use of these devices both for hybrid and optical applications.

## 1 Introduction

Gapped metallic nanorods provide an effective way to concentrate and manipulate light at the nanoscale through the excitation of collective charge oscillations known as surface plasmons.[1,2] When working at a fixed wavelength of the incoming light the geometrical parameters of the sample can be optimized to produce localized surface plasmon resonances (LSPRs) leading to the enhancement of the electric field inside the nanogap.[3] Due to this hot-spot effect these nanostructures can be used as antennas for the efficient collection of light, in analogy to what is done in the area of microwaves and radiofrequencies.[4–6] This behavior has attracted a great deal of interest in the last decade due to its wide range of applications in nanotechnology, such as signal amplification,[7] molecular recognition, and biosensing.[8,9] This is achieved through a wide variety of optical sensing and spectroscopy techniques, most notably surface-enhanced Raman spectroscopy (SERS),[7,10–12] giving access to detect particular vibrations identifying target molecules lying in the gap area. The figure of merit characterising the nanoantenna efficiency for SERS spectroscopy is the well known enhancement factor (EF) which integrates the fourth power of the intensity of the electric field over a surface inside the gap (see eqn (1)).[13,14] Crucially, even a slight increment in the EF can prevent false-positive and false-negative results leading to an overall improvement of biosensing and biomarker applications.[15,16]

In previous works, the effects of the geometrical parameters of gold gapped nanorods on the EF at the gap were studied theoretically and experimentally.[17] The SERS intensity at the gap of these structures has a periodic dependence on the length of the gold segment in agreement with EF calculations. The interaction between light and surface plasmons generates surface plasmon polaritons (SPPs) traveling along the cylinder. As Li *et al.*[18] pointed out, for fixed nanorod diameter and length, the SPPs arrange in allowed standing waves. In particular, the EF peaks are associated with LSPRs arising when the standing wave contains odd-symmetry SPP modes in each segment. Importantly, the dimer antenna performance is good even at resonances associated to a several SPP wavelengths allowing the nanorod lengths to be larger than a few micrometers. Cylindrical gapped gold nanoantennas have been fabricated by onwire lithography.[19–22] Bowtie type or rectangular nanoantennas have been synthesized by electron beam microscopy.[6,23,24]

As the enhancement factor can reach peak values with arm lengths surpassing the micrometer, an important feature of these structures is that they can be used as addressable electrodes.[25,26] Electrical transport measurements can be coupled with the ability to spectroscopically characterize target molecules at the gap by SERS, either sequentially,[27,28] or simultaneously, as would be our final goal. In order to use these structures as electrodes the rods must be contacted,[29] *i.e.*, a certain amount of gold must be deposited on the nanorods and this could affect its plasmonic properties.

[a]Facultad de Ciencias, Universidad Nacional de Ingenieria, Apartado 31-139, Av. Túpac Amaru 210, Lima, Perú. E-mail: mleon@uni.edu.pe

[b]Centro Atómico Bariloche & Instituto Balseiro, Instituto de Nanociencia y Nanotecnología, CNEA-CONICET, Av. Bustillo 9500, Bariloche, Argentina. E-mail: reynoso@cab.cnea.gov.ar

Some authors have used an approximated method to simulate Fabry–Perot resonances in a single cylindrical wire;[30] FDTD simulations to study the near-field optical behavior of Fabry–Perot resonances in thin metal nanowires referred to as quasi one-dimensional plasmonic nanoantennas;[31] characteristic mode analysis to optimize complex plasmonic nanoantennas;[32] integral equations formulation solved using the Method of Moments to study a plasmonic wire-grid array of nanorods;[33] the commercial finite-difference time-domain (FDTD) software (Lumerical FDTD) to study the electric near-field enhancement of cascaded plasmonic nanorod antenna;[34] and a circuit equivalent of a plasmonic nanoantenna.[35]

However, to the best of our knowledge, until now no theoretical and experimental study has been reported about the geometrical effects on the EF when these nanorods are covered with gold. In this work, using a discrete dipole approximation (DDA) method,[36–38] we theoretically study the behavior of the enhancement factor in symmetric gapped nanorods subject to geometrically designed gold coverings at the far-from-the-gap regions of the dimer.

The paper is organized as follows. In Section 2 we introduce our covered nanoantenna designs and the potential approaches for their fabrication, describe the applied numerical method, and define several relevant quantities as the enhancement factor (EF) and the top and bottom average surface charge densities. The later location-resolved densities, introduced for the first time here, prove to be useful in the qualitative understanding of the improvement of the EF as a result of a geometrically-assisted in-phase rearrangement of the top and bottom SPPs. In Section 3 we present the simulations for covered nanoantennas demonstrating that, even in the presence of dimer asymmetries or defects, the reached EF values surpass the highest values obtained in uncovered nanorod dimers. Finally, in Section 4 we present the conclusions.

## 2 Covered nanoantenna: designs and modeling

### Geometry and optical setup

The proposed covered designs use as a starting point the dimer configuration of two cylindrical gold nanorods separated by a small gap, as shown in Fig. 1 (top). All along this work, envisioning applications of spectroscopic detection of molecules such as DNA, we consider a gap length of 25 nm. We also fix the diameter of the cylinders in 360 nm since this case can be synthesized with good quality and excellent optical properties as demonstrated in ref. 18. The available fabrication methods ensure that the two nanorods are well aligned (we take this as the *z* axis) achieving large EF within the gap when the dimensions are properly chosen,[17,18] see Fig. 2 below. In what follows we refer to this initial design as the uncovered nanoantenna. Following the customary optical setup of ref. 18, we considered the situation in which the nanoantenna is subject to an excitation from a 633 nm wavelength helium–neon continuous-wave laser at normal incidence. The incoming wave is polarized along *z* in this way it maximizes the field in the gap area. The main qualitative conclusions presented below can be straightforwardly extended to other values of diameter, gap and incoming wavelength.

Selective deposition of gold allows for the syntheses of samples in which each arm of the dimer has a covered region of length *C* located at the far-from-the-gap region. As shown in the sketches for designs A and B in Fig. 1, each arm has a total length of $L = C + U$ where $U$ is the length of the uncovered near-gap region. The deposited gold layer has height *h*, this quantity enlarges the diameter of the circular section describing the covered part of each arm. Such circle is taken tangential to the plane of the substrate ($x = 0$) in order to account for the fact that the gold-evaporation process is more efficient at the top of the cylinders. The design B has an extra step of fabrication in which gold at both arms is etched vertically (planes of *y* constant, see

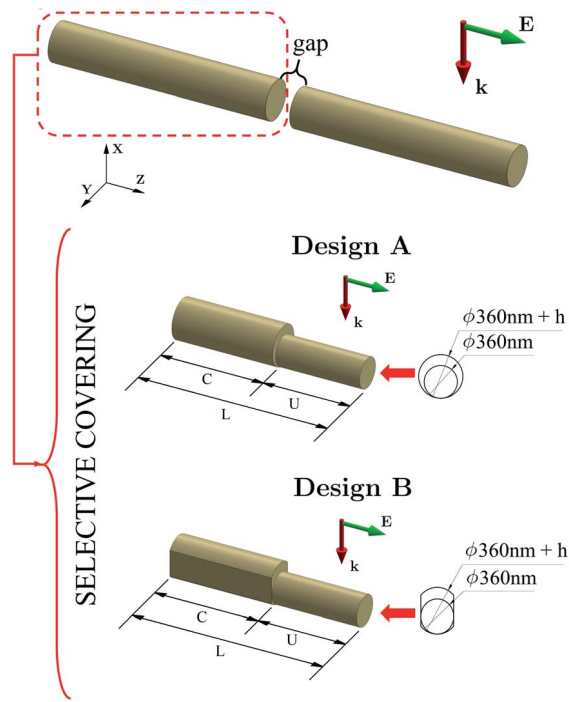

**Fig. 1** The starting design is the uncovered nanorod dimer (top) which is known to have good enhancement factor enabling SERS spectroscopy on target molecules positioned in the gap region. We use feasible system dimensions and customary optical setup (see for example ref. 18) having a gap length of 25 nm and subject to $\lambda = 633$ nm normal incident light polarized along *z* axis, *i.e.*, parallel to the axis of the cylinder, thus maximizing the electric field enhancement in the gap. For the prospect of electrically contacting the system without degrading the enhancement factor our goal is studying the optical response after covering the outer regions of the rods leaving the gap region unaltered. We focus on two geometrical implementations, design A: a cylindrical cover, and design B: a cylindrical cover laterally etched. For both cases we show the 3-dimensional shape of the nanorods and the definition of the geometric parameters (for clarity only one dimer's arm is shown).

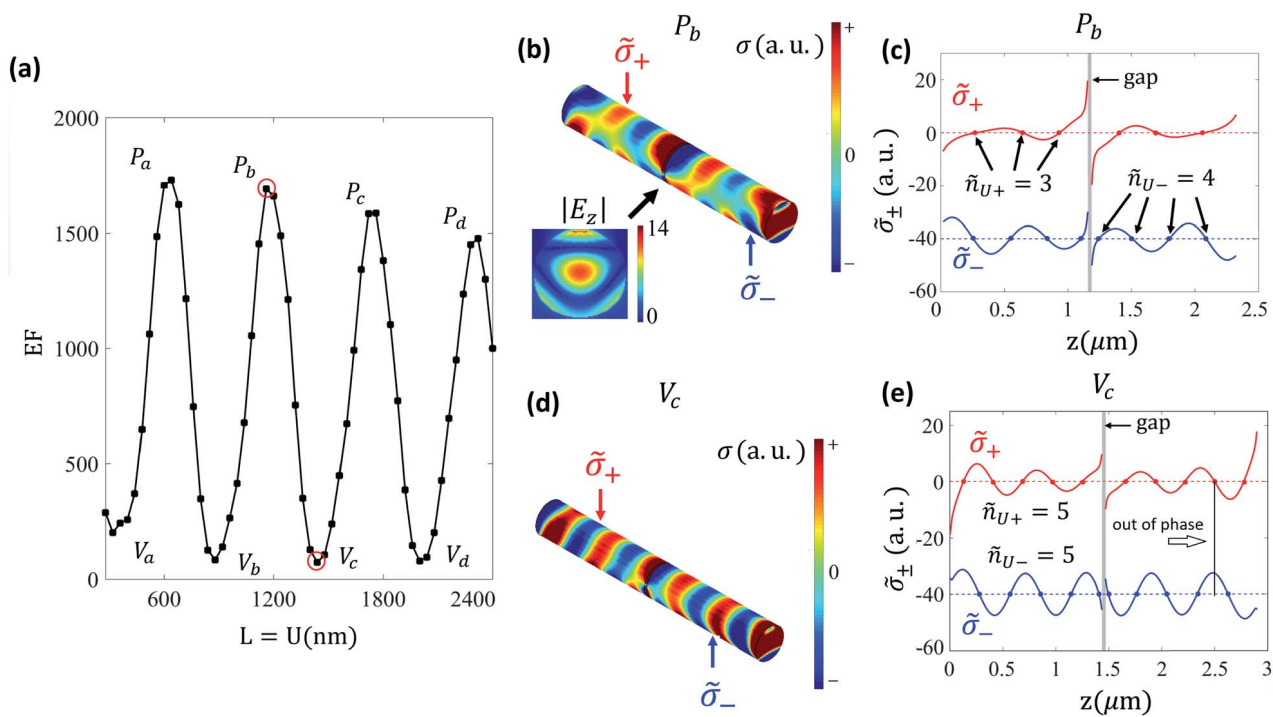

**Fig. 2** Simulated optical response of the uncovered nanorod-dimer antenna design of Fig. 1 (top). (a) Enhancement factor *versus* nanorod length $L$. For specific values of $L$ the EF has peaks and valleys. (b) and (d) Surface charge density on the dimer antenna for peak $P_b$ (valley $V_c$). For quantitatively visualizing the top and bottom differences panels (c) and (e) present the top, $\tilde{\sigma}_+(z)$, and bottom, $\tilde{\sigma}_-(z)$, average surface charge densities for peak $P_b$ (valley $V_c$). A colormap of $|E_z|$ at the gap for peak $P_b$ is included as an inset of panel (b). Notice that at the peak condition, $\tilde{n}_{U+} \neq \tilde{n}_{U-}$, *i.e.*, the number of nodes in the top and bottom average charge densities are different. This top and bottom mismatch also appears in all EF peaks, see Table 1. At valley conditions, although $\tilde{n}_{U+} = \tilde{n}_{U-}$, the top and bottom surface charge averages are out of phase.

Fig. 1) restricting the covered parts to have dimension along $y$ equal to the uncovered diameter. In this way, the design B has the advantage of being unaffected by potential lateral imperfections of the covered parts.

It is well known that for fixed wavelength, diameter, and gap distance, the uncovered dimer configuration of ref. 17 and 18 must be fabricated at particular nanorod lengths to ensure working at a local maximum of EF. In these works, the numerical simulations of the enhancement factor we are following, see eqn (1) below, lead to excellent agreement with the measured SERS of target molecules lying in the gap. One could do numerical simulations fixing the geometry and finding the optimal wavelength that maximizes the EF. Instead, as in ref. 17 and 18, simulations fixing the wavelength and diameter provide more practical information for experimental purposes—assuming a standard laboratory having 633 nm wavelength helium–neon sources and sample templates of fixed diameters, see next section. The EF optimization is performed by controlling geometrical parameters that are easier to change in the design as it is the length of the nanorods.

### Different fabrication approaches

On the one hand, the chemical synthesis of the metallic cylinders can proceed by means of well-established methods for the formation of colloidal nanorods[39] or nanowires.[40,41] Electrochemical on wire-lithography (OWL) methods using inorganic porous scaffolds are also available.[17] Subsequently, the particles would be deposited on a substrate and submitted to different types of nanostructuring by means of top-down techniques, including electron beam lithography and focused ion beam milling. One possibility is to cover the deposited nanorods or nanowires with a positive-tone resist and perform electron beam lithography to uncover part of the cylinders.[42] Metal deposition on the exposed regions, by either electrodes metal reduction or by metal evaporation, would complete the additional diameters for the part C in the designed structure followed by lift-off. A similar process could be used to open the gap region through photo-resist covering, electron beam lithography to uncover the gap and finally the metal etching procedure. An alternative route to both opening the gap and etching the lateral faces of region C is by focused ion beam, in which a focused beam of Ga$^+$ ions is used for the removal of material with nanometric precision.[43,44] In summary, there exists a variety of available techniques that could be applied in the synthesis of high quality design A and B or similar covered nanoantennas.

In particular, the OWL technique, used in ref. 17 and 18, is described in ref. 22 achieving sub-5 nm fabrication resolution. One typically fixes the nanorod diameter to be grown by electrochemical deposition of metals within the pores of anodic aluminum oxide (AAO) templates. We stress that the diameter value of 360 nm has proven to be an excellent choice for the AAO templates; nanorods of smaller diameters tend to break, thus defeating the purpose of obtaining long arm dimers that could be covered and electrically contacted, as it is our goal here. This, again, justifies the fixing of the diameter in our study to be consistent with the feasible value fabricated and simulated in ref. 17 and 18. Reproducing these known results for uncovered dimers ensures a solid ground from which one can evaluate and qualify the effect of the covering modifications. Importantly, as our covered design introduces modifications in the regions far from the gap, the procedure to compute the EF for predicting the SERS performance remains unchanged and identical to these relevant references in the field.

### Applied numerical method

In order to solve the Maxwell equations and obtain the electric field, we make use of the discrete-dipole approximation (DDA) method,[45] which has been successfully applied to the uncovered nanoantenna design.[18] In particular, we use the well established and robust DDSCAT 7.3 implementation of the DDA method.[36,37]

The input parameters (coordinate of the dipoles and material properties) were generated for each nanoantenna design using a custom-made FORTRAN code. The dipoles are arranged in a cubic lattice (with lattice spacing here taken 5 nm) suitable for an speed-up *via* the fast Fourier transform as implemented in the DDSCAT package.[38] As we work with a frequency of $\omega = 2\pi c/(633$ nm) the real and imaginary part of gold's dielectric constant are $\varepsilon_1 = -11.79916$ and $\varepsilon_2 = 1.22127$, respectively. Once the full solution is obtained, we post-process the output and obtain the electric field at the surface and the gap area of the nanoantenna and proceed to compute the quantities defined in the next section.

### Enhancement factor and surface charge densities

In order to characterize the efficiency for SERS applications of the antennas we use as a figure of merit the enhancement factor,[13,14] $g^4 = |\mathbf{E}|^4/|E_0|^4$, where $\mathbf{E}$ is the local electric field and $E_0$ is the amplitude of the incident electric field. This enhancement is averaged over a surface at the gap region:

$$\mathrm{EF} = \frac{\int |\mathbf{E}|^4/|E_0|^4 \mathrm{d}S}{\int \mathrm{d}S}, \tag{1}$$

where d$S$ is the surface area differential over a plane parallel to one of the rod faces locate 2.5 nm inside the gap. We stress that for the purpose of predicting an efficient nanoantenna design, in practice, it suffices computing the EF at a single surface as it is done here. In the particular case of gapped nanorod dimers, Pedano *et al.*[17] and Li *et al.*[18] have shown that the experimentally measured SERS signal is well described by the EF computed at the position that we are adopting here. Notice that the covered designs, for the prospect of electrical contacting the nanoantenna, involve modifications in regions far from the gap. Therefore, for quantifying the ability to detect target molecules inside the gap with SERS, it is well justified the adoption of the experimentally validated method to compute the EF presented in the mentioned relevant references in the field of gapped gold nanorod dimers. In the next section, we start from solid grounds by reproducing the EF obtained in uncovered designs and proceed to evaluate the effect of the covering modifications.

It is useful to have an additional way to characterize a given design, specially for geometrical conditions in which the EF achieves a maximum or minimum. In these extremal conditions we resort to analyse the surface charge density of the antenna: the patterns of changes of sign of the surface charge along the device, allow for identifying the morphology of the involved plasmonic excitations. Even though the covered designs have a complex geometry, sometimes the parity of the excitations can be identified.

For obtaining the surface charge density $\sigma(\mathbf{r}_s)$ we apply, at each position lying on the surface, $\mathbf{r}_s$, the Gauss law $\varepsilon_0(\mathbf{E}_{\mathrm{out}}(\mathbf{r}_s) - \mathbf{E}_{\mathrm{in}}(\mathbf{r}_s)) \cdot \hat{\mathbf{n}}(\mathbf{r}_s) = \sigma(\mathbf{r}_s)$: where $\mathbf{E}_{\mathrm{out}}(\mathbf{r}_s)$ ($\mathbf{E}_{\mathrm{in}}(\mathbf{r}_s)$) is the electric field outside (inside) the metal and $\hat{\mathbf{n}}(\mathbf{r}_s)$ the unit normal to the sample at that position. As our goal is the qualitative inspection of the plasmonic modes, we work with an approximated charge density obtained using the numerically obtained $\mathbf{E}_{\mathrm{out}}(\mathbf{r}_s)$ and collapsing all the charge to a thin sheet by assuming the perfect conductor limit ($\mathbf{E}_{\mathrm{in}}(\mathbf{r}_s) = 0$). This procedure suffices to capture the patterns of charge density induced by the plasmonic modes of interest.

For a simpler characterization of the plasmonic $z$-dependence along each cylindrical uncovered part of the nanoantenna, we perform, at each $z$, the average of the surface charge density along the circle. Furthermore, we divide this average into the top and bottom part of the nanoantenna: as we show below, this is useful for distinguishing differences between peak and valley conditions and how they change in the covered designs. The definition of these top and bottom surface charge density averages are given by:

$$\tilde{\sigma}_+ (z) = \pi^{-1} \int_{-\pi/2}^{\pi/2} \mathrm{d}\phi \sigma(\phi, z) \tag{2a}$$

$$\tilde{\sigma}_- (z) = \pi^{-1} \int_{\pi/2}^{3\pi/2} \mathrm{d}\phi \sigma(\phi, z) \tag{2b}$$

It is particularly useful to account for the number of sign changes along $z$ of the former quantities. These number can be written as

$$\tilde{n}_{U\pm} = \int_{L-U}^{L} \mathrm{d}z \left| \frac{\mathrm{d}\tilde{\sigma}_\pm(z)}{\mathrm{d}z} \right| \delta\left(z - \tilde{\sigma}_\pm(z)\right) \tag{3}$$

## 3 Results

### Uncovered design

We start by computing the EF for the uncovered nanoantenna design of Fig. 1(top). Fig. 2(a) presents the behavior of EF as a function $L$, or equivalently $U$, since here $C = 0$. As it is well known,[18] a successful design must be fabricated having special values of $L$, thus achieving EF of the order of 1700. We label these peak conditions as $P_a$, $P_b$, $P_c$, and $P_d$. (Similarly we label the valley conditions $V_a$, $V_b$, $V_c$, and $V_d$ since it is conceptually important exploring these situations.) As shown below the geometric modifications involved in the covered designs generate EF values even larger than the best cases achieved in the uncovered design.

**Table 1** Enhancement factor, $\bar{n}_{U+}$ and $\bar{n}_{U-}$ for the uncovered configuration (here, since $C = 0$, $U = L$) at peaks and valleys: see labels in Fig. 2. Notice the opposite sign of $\bar{\sigma}_+(z)$ and $\bar{\sigma}_-(z)$ at the positions $z_e$: these are the locations of both external edges of each of the dimer's arm, namely, $z_e = 0$ and $z_e = 2L + 25$ nm.

|       | $L$ [nm] | EF   | $\bar{n}_{U+}$ | $\bar{n}_{U-}$ | sgn($\bar{\sigma}_+(z_e)\bar{\sigma}_-(z_e)$) |
|-------|----------|------|----------------|----------------|-----------------------------------------------|
| $V_a$ | 320      | 202  | 1              | 1              | $-1$                                          |
| $P_a$ | 640      | 1729 | 1              | 2              | $-1$                                          |
| $V_b$ | 880      | 84   | 3              | 3              | $-1$                                          |
| $P_b$ | 1160     | 1691 | 3              | 4              | $-1$                                          |
| $V_c$ | 1440     | 73   | 5              | 5              | $-1$                                          |
| $P_c$ | 1760     | 1586 | 5              | 6              | $-1$                                          |
| $V_d$ | 2000     | 79   | 7              | 7              | $-1$                                          |
| $P_d$ | 2320     | 1476 | 7              | 8              | $-1$                                          |

Fig. 2(b) shows the pattern of $|E_z|$ at the gap and the surface charge charge density for the uncovered antenna of $L = 1160$ nm, which is the case of the peak $P_b$ with EF = 1691. It is seen that $\sigma(\mathbf{r}_s)$ has a different structure at the top and the bottom (with respect to the $x$ coordinate) of the nanoantenna. This difference becomes evident in Fig. 2(c) where $\bar{\sigma}_-(z)$ and $\bar{\sigma}_+(z)$ are plotted. Moreover, the number of nodes (positions in which the average charge densities change sign) is $\bar{n}_{U+} = 3$ (i.e., odd) and $\bar{n}_{U-} = 4$ (i.e., even) for $\bar{\sigma}_+(z)$ and $\bar{\sigma}_-(z)$, respectively. The same quantities are plotted in Fig. 2(d–e) for a nanoantenna of $L = 1440$ nm corresponding to the valley $V_c$ with EF = 73. In this case there is an out of phase relation between $\bar{\sigma}_+(z)$ and $\bar{\sigma}_-(z)$ and, at the same time, $\bar{n}_{U+} = \bar{n}_{U-} = 5$ (i.e., odd). From the oscillating $z$-dependence of the charge densities one can estimate the SPP wavelength arising from the simulation to be near $\lambda_p = 600$ nm, in close agreement with previous reports for a gap of $24\text{nm}$.[18]

In Table 1 we list the obtained values of $\bar{n}_{U\pm}$ for the peaks and valleys of EF. In the case of $\bar{n}_{U+}$ it is always odd, increasing with $L$ every two EF extrema; while $\bar{n}_{U-}$ takes all positive integer values. We find that in the EF valleys $\bar{n}_{U+} = \bar{n}_{U-}$ and odd. On the other hand, in the EF peaks $\bar{n}_{U-} = \bar{n}_{U+} + 1$, meaning that $\bar{n}_{U-}$ is even whereas $\bar{n}_{U+}$ is odd. The odd parity of $\bar{n}_{U+}$ is consistent with the interpretation that from the resonances of the single cylinder system, arising when $L = (n - 1/2)\lambda_p/2$,[30,46] only the odd ones survive for incoming light polarized along the cylinder axis. The conditions in the length of the dimer arms for obtaining the resonances get modified when two cylinders couple and interact in a way that depends on the gap distance and cylinder diameter.[18] Table 1 also highlights that for both valleys and peaks the averages densities $\bar{\sigma}_\pm(z)$ have opposite signs at the external positions of the nanorods (see examples in Fig. 2(c) and (e)) $z_e = 0$ and $z_e = 2L + 25$ nm, and thus sgn($\bar{\sigma}_+(z_e)\bar{\sigma}_-(z_e)$) = $-1$. These observations—that are particular to the current ratio between the SPP wavelength and circular perimeter of the nanorods—are useful in the comparison to the covered designs because they encode qualitative and quantitative features of the SPP generated surface charge patterns at the extrema of EF.

## Improvement of EF in the covered designs

To analyze how EF changes due to the geometrical modifications imposed by the gold covering, we first study the effect of the gold layer height $h$ for the covered designs introduced in Fig. 1. Fig. 3 shows the nanoantennas' EF versus $L$ for different values of $h$, while keeping $C = 1200$ nm. This $C$ is very close to twice the SPP wavelength which below is shown to be an optimal design choice. Changing the total length in this case implies enlarging $U$ since $U = L - 1200$ nm. We can see that the increase of the parameter $h$ shifts the position of the peaks while the EF maximum grows. Both cases show a great improvement of the enhancement factor compared to the case without cover ($h = 0$), obtaining a maximal EF equal to 4378 at ($h = 120$ nm, $L = 1680$ nm) for the design A, and EF equal to 5242 at ($h = 320$ nm, $L = 2160$ nm) for the design B.

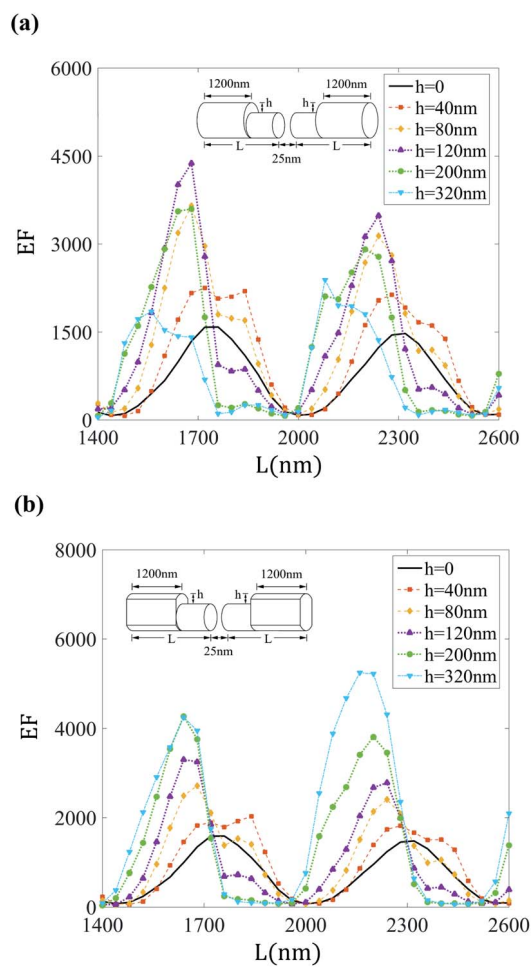

**Fig. 3** Enhancement factor versus length $L$ for different values of $h$ taking $C = 1200$ nm in the covered designs shown in Fig. 1. In both designs the added covering layer, even though it is added far from the gap, can generate larger EF than in the uncovered design of $h = 0$. The position of the largest obtained EF values are similar in both designs.

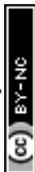

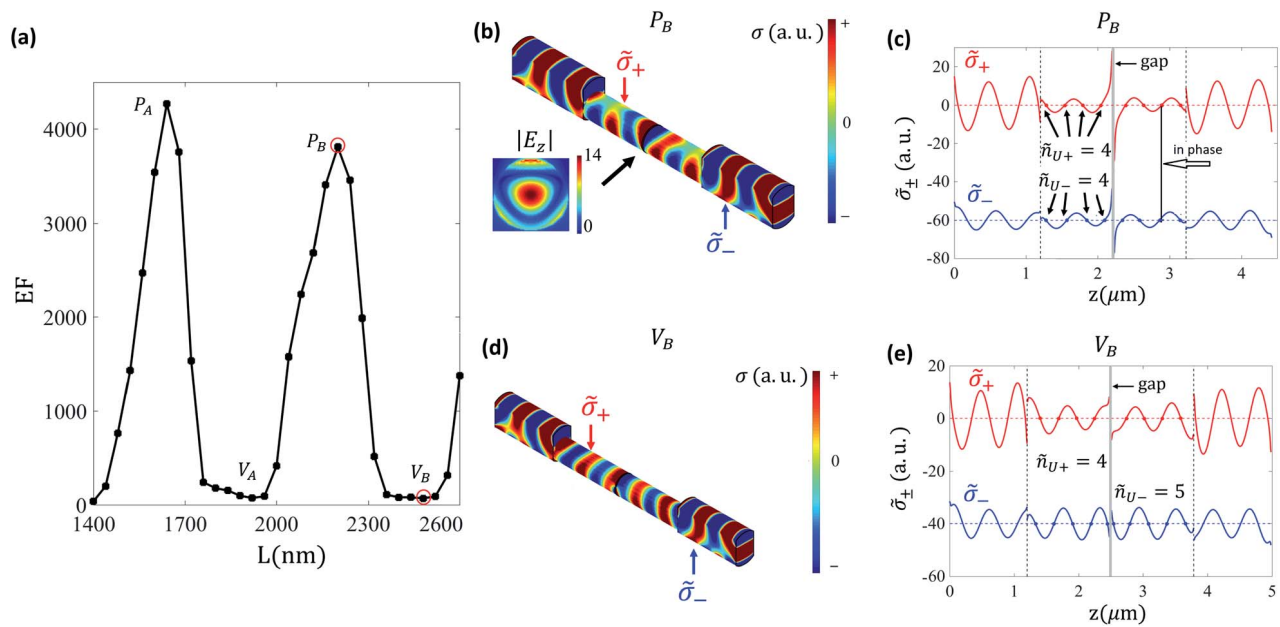

**Fig. 4** Simulated optical response of the covered design B presented in Fig. 1. (a) Enhancement factor *versus* nanorod length *L*. For specific values of *L* the EF has peaks and valleys. (b) and (d) Surface charge density on the dimer antenna for peak $P_B$ (valley $V_B$). For quantitatively visualizing the top and bottom differences panels (c) and (e) present the top, $\tilde{\sigma}_+(z)$, and bottom, $\tilde{\sigma}_-(z)$, average surface charge densities for peak $P_B$ (valley $V_B$). A colormap of $|E_z|$ at the gap for peak $P_B$ is included as an inset of panel (b). Notice that for the peak condition, in contrast to the uncovered case presented in Fig. 2, here $\tilde{n}_{U+} = \tilde{n}_{U-}$, *i.e.*, the number of nodes in the top and bottom average charge densities are equal (see Table 2). This top and bottom in phase matching (see panel(c)), induced by the geometrical modification far from the gap, allows the covered design to achieve larger EF values than the uncovered dimer.

Our simulations demonstrate that the covered designs, if they are built with the appropriate geometrical parameters, not only do not degrade the EF but also achieve values of EF that double or even triple the EF value obtained in an optimal designed uncovered nanorod dimer. This is one of the main results of the paper: that this geometry-induced EF improvement appears in slightly different designs. We note at this point that, due to its lateral etching, the design B is more robust to imperfections of the gold layer produced in the covering process. For this reason, in what follows, we explore in detail the case of design B. Design A behaves qualitatively similar than design B, *i.e.*, it achieves large EF values sharing the below presented robustness to imperfections.

### Surface charge densities in peaks and valleys

Fig. 4(a) shows the improved EF as function of *L* for the design B with $C = 1200$ nm and $h = 200$ nm. In Fig. 4(b) and (d), for the *L* value of the peak $P_B$ and the valley $V_B$ (see labels in Fig. 4(a)), the surface charge density is shown using 3D positioned colormaps. We also present the corresponding top and bottom average charge densities, see eqn (2), as a function of *z* in panels (c) and (e). It is important to note that, as shown in the inset of Fig. 4(b), the $|E_z|$ colormap at the gap corresponding to peak $P_B$ has the same mode than the uncovered design (see the inset in Fig. 2(b)) but with larger amplitude. Therefore the reason of this enhancement, as we discuss below in more detail, is not related to intra-gap physics.

Table 2 presents relevant data for a sequence of peaks and valleys corresponding to both the $C = 1200$ nm case of Fig. 4

and the $C = 120$ nm case (not shown). A first difference from the uncovered case is that the top and bottom average charge densities have the same sign at the edges of the antenna, *i.e.*, $\mathrm{sgn}(\tilde{\sigma}_+(z_e)\tilde{\sigma}_-(z_e)) = +1$. The covering also modifies the behaviour of the numbers of nodes of the average densities along the uncovered parts, *U*, see eqn (3). We find that at the EF peaks $\tilde{n}_{U+} = \tilde{n}_{U-}$ and even: see for example the case shown in Fig. 4(c) with $\tilde{n}_{U\pm} = 4$, having an in phase relation between $\tilde{\sigma}_+(z)$ and $\tilde{\sigma}_+(z)$ along the uncovered part. This is in contrast to what happens in the uncovered design of Fig. 2, see Table 2, in which: (i) at EF-peak conditions there is a parity mismatch between the top and bottom average charge densities, and (ii) at valley conditions one finds odd $\tilde{n}_{U+} = \tilde{n}_{U-}$ but having an out of phase relation. Therefore, for designs with optimal *C* values (see

**Table 2** EF, $\tilde{n}_+$ and $\tilde{n}_-$ for design B covered nanoantennas of $h = 200$ nm with either $C = 120$ nm or $C = 1200$ nm at *U* lengths developing peaks or valleys. The labels for the $C = 1200$ nm case are presented in Fig. 4(a)

|  | $(C, U)$ [nm] | EF | $\tilde{n}_{U+}$ | $\tilde{n}_{U-}$ | $\mathrm{sgn}(\tilde{\sigma}_+(z_e)\tilde{\sigma}_-(z_e))$ |
|---|---|---|---|---|---|
|  | (120, 440) | 4628 | 2 | 2 | +1 |
|  | (120, 720) | 66 | 2 | 3 | +1 |
|  | (120, 1000) | 3937 | 4 | 4 | +1 |
|  | (120, 1280) | 71 | 4 | 5 | +1 |
| $P_A$ | (1200, 440) | 4271 | 2 | 2 | +1 |
| $V_A$ | (1200, 720) | 74 | 2 | 3 | +1 |
| $P_B$ | (1200, 1000) | 3810 | 4 | 4 | +1 |
| $V_B$ | (1200, 1280) | 70 | 4 | 5 | +1 |

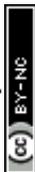

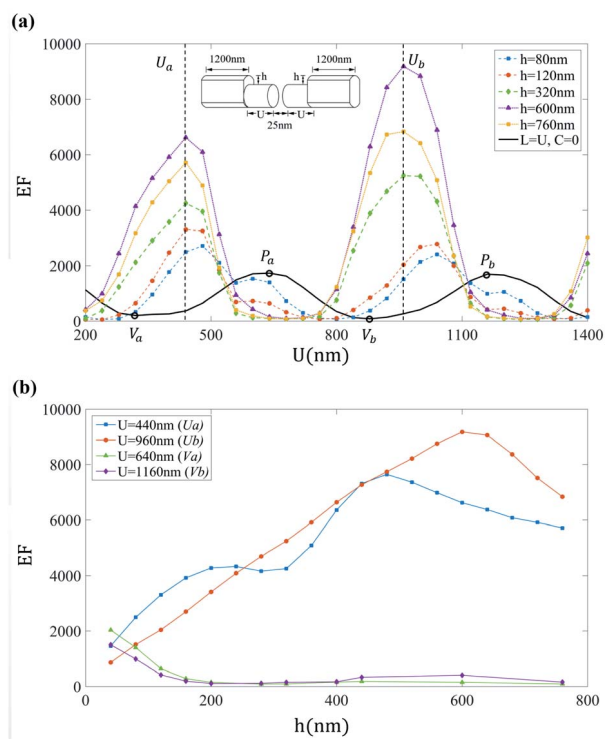

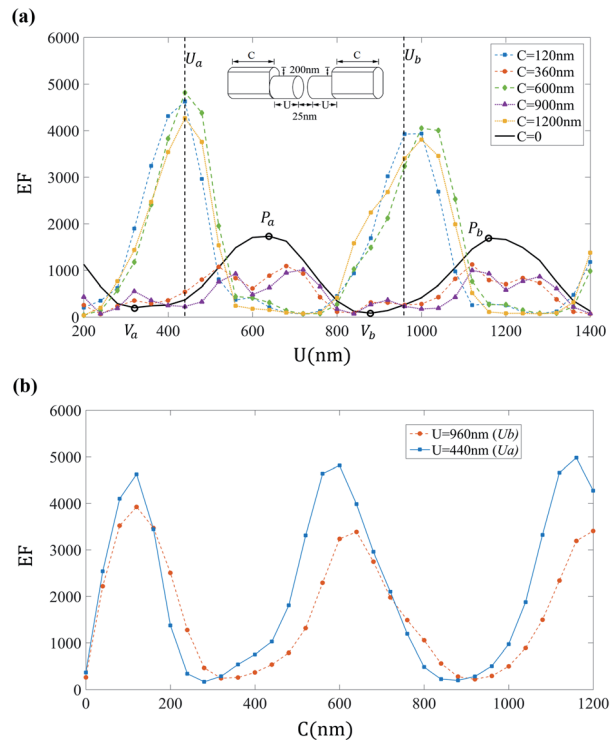

**Fig. 5** (a) Enhancement factor *versus* uncovered length $U$ for different values of $h$ in the covered design B with $C = 1200$ nm. The dotted lines, labeled by $U_a$ and $U_b$, are located at the $U$ values in which the EF peaks develop. Notice they lie near the position of the EF valleys for an uncovered antenna with full length $U$. Moreover, panel (b) shows that for values of $U$ that produce peaks in an uncovered antenna of full length $U$, the covered nanoantenna of length $L = U + 1200$ nm worsens its EF when $h$ grows. Thus, a good design choice based solely on the knowledge of the uncovered case is choosing an uncovered length that would generate a valley if that section were an isolated uncovered dimer.

**Fig. 6** (a) Enhancement factor *versus* uncovered length $U$ for different values of $C$ in the covered design B with $h = 200$ nm. Notice that, once again, the obtained EF is not improved by the existence of a $C$ part for $U$ values that would develop a peak (see $P_a$ and $P_b$) if the uncovered section were an isolated dimer. The dotted lines, labeled by $U_a$ and $U_b$, are the ones defined in Fig. 5 lying near the position of the EF valleys for an uncovered antenna with full length $U$. Panel (b) shows the EF dependence as a function of $C$ at $U_a$ and $U_b$. These values continue being favourable choices for achieving large enhancement.

below), the geometrical modification imposed by the addition of the covered part makes the top and bottom average charge densities in the $U$ region to become in phase, thus enabling the possibility of achieving higher EF values than the largest EFs obtained with the uncovered design.

### Dependence with geometrical parameters

Fig. 5(a) shows the EF values as a function of the uncovered length $U$ of the design B with $C = 1200$ nm for different covering gold height $h$. We only focus on $h$ values up to the order of the diameter of the original cylinder (360 nm) since larger values of $h$ makes the sample to differ substantially from the original geometrical configuration, thus falling beyond the scope of this study. Fig. 5(a) also presents the EF curve for the fully uncovered case having each arm length equal to $U$ (note this corresponds to the same curve given in Fig. 2(a)). In Fig. 5(b) we plot the EF dependence with $h$ for fixed values of $U$. We find that the choice of $U$ values near peak conditions in the uncovered design are undesirable as they lead to low values of EF. On the other hand, the EF grows with $h$ up to 600 nm when choosing $U$ values near the valley conditions of the uncovered design: notice they fall near the $U_a$ and $U_b$ conditions in which the largest EF values are achieved. This is consistent with the fact that both at EF valleys of the uncovered design and at EF peaks of the covered design it holds that $\tilde{n}_{U^+} = \tilde{n}_{U^-}$.

For the covered design B with $h = 200$ nm Fig. 6(a) shows the enhancement factor as function of the uncovered length $U$ for different covered lengths $C$. For $U$ values in which an uncovered dimer with arms of length $U$ would develop a peak (see $P_a$ and $P_b$) the enhancement factors of the covered design is small. Fig. 6(b) shows the EF dependence as a function of $C$ for the $U_a$ and $U_b$ values presented in Fig. 5: the ones lying near the position of the EF valleys for an uncovered antenna with full length $U$. As in the case of Fig. 5 this choices of $U$ values are good candidates for amplifying the EF with respect to the uncovered design. In particular large EF values are reached at a broad region near $C$ lengths multiples of the surface plasmon polariton wavelength, *e.g.* 600 nm and 1200 nm. However, very low EF values are obtained for $C$ lengths near odd multiples of $\lambda_p/2$, *e.g.* 300 nm and 900 nm and thus they must be avoided.

### Robustness to geometrical asymmetries and defects

Given a particular sample fabrication and design processes, the detailed geometry and quality of the dimer can be very different requiring the realization of sample specific simulations.

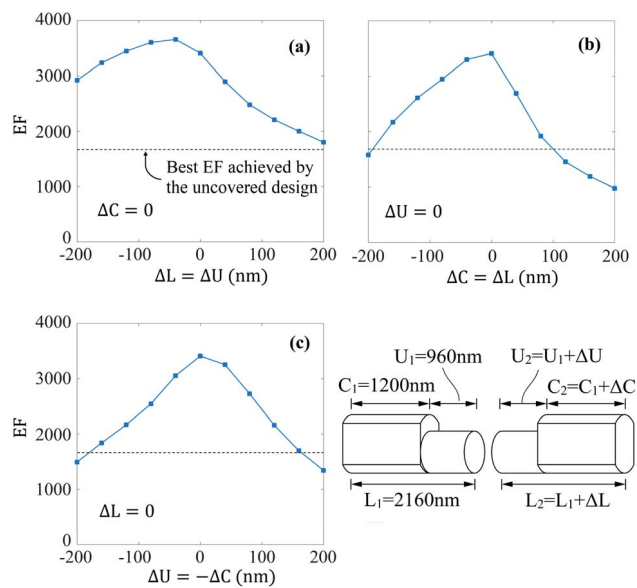

Covering all possible cases falls beyond the scope of this work, however, it is important to explore some simple scenarios of sample imperfections. We aim at knowing how the improvement of the enhancement factor, demonstrated above for the case of a perfect dimer, behaves in the presence of geometrical asymmetries in the fabrication. For this, as shown in the bottom right panel of Fig. 7, we consider a dimer in which the first arm has total length $L_1 = C_1 + U_1$ with $C_1$ ($U_1$) its (un)covered length and the second arm has total length $L_2 = C_2 + U_2$ with $C_2$ ($U_2$) its (un)covered length. We define the differences $\Delta L = L_2 - L_1$, $\Delta C = C_2 - C_1$, and $\Delta U = U_2 - U_1$.

In Fig. 7 we present simulations of the enhancement factor for three cases of geometrical asymmetries in the design B with $h = 200$ nm. The first dimer arm is always fixed at $U_1 = 960$ nm and $C_1 = 1200$ nm so that the obtained EF value would be 3405 in the case of perfect symmetry, $\Delta C = \Delta U = \Delta L = 0$. This EF value is slightly below the EF at the peak $P_B$ presented in Fig. 4(a) and listed in Table 2. Similar quantitative conclusions can be reached by working with the first arm dimensions ($C_1$, $U_1$) set to other values in which perfect symmetry would make the dimer to reach an enhancement factor peak surpassing the uncovered nanoantenna performance. In Fig. 7(a) we consider the case of $\Delta C = 0$ and show the EF dependence with $\Delta U = \Delta L$. In Fig. 7(b) we consider the case of $\Delta U = 0$ showing the EF dependence with $\Delta C = \Delta L$. These two simulations show that, in case of good matching one of the lengths (e.g., either $U_2$ or $C_2$ is well matched to the first arm), it is better having negative fabrications errors in the other length (i.e., $U_2 < U_1$ or $C_2 < C_1$). Fig. 7(c) shows the EF dependence with $\Delta U = -\Delta C$ for the case of $\Delta L = 0$. Notably, all these simulations show that, even for 100 nm errors, the reached enhancement factor still surpasses the one obtained at the EF peaks of the uncovered design.

**Fig. 7** Robustness of the enhancement factor to differences in the lengths between the first and second arm of the dimer: see sketch at the bottom right panel defining $\Delta L = L_2 - L_1$, $\Delta U = U_2 - U_1$, and $\Delta C = C_2 - C_1$. This design B dimer has a $h = 200$ nm gold layer and the first arm is taken along the line $U_b$, see Fig. 5(a), with $U_1 = 960$ nm and $C_1 = 1200$ nm: if the second arm were identical the reached EF would be 3405, i.e., slightly below the value at the peak $P_B$ given in Table 2. (a) Enhancement factor for $\Delta C = 0$ as a function of $\Delta U = \Delta L$. (b) Enhancement factor for $\Delta U = 0$ as a function of $\Delta C = \Delta L$. (c) Enhancement factor for $\Delta L = 0$ as a function of $\Delta U = -\Delta C$. In all cases the error can be of the order of 100 nm and the resulting EF is still higher than the best obtained EF value in the uncovered design: see the black dotted line.

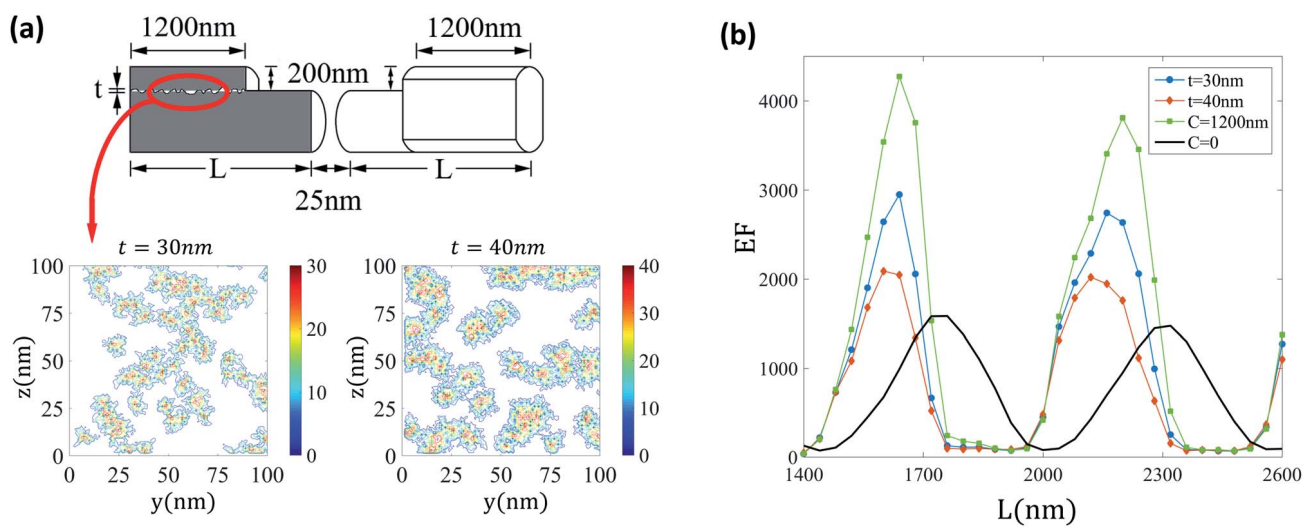

**Fig. 8** Robustness of the enhancement factor to defects located at the interface between the original cylinder and the covering layer. (a) Top: Sketch locating the thickness $t$ layer in which vacancies are introduced. Bottom: Colormaps showing portions of the randomly generated patterns of vacancies corresponding to $t = 30$ nm and $t = 40$ nm. (b) Enhancement factor as a function of $L$ for the two situations with defects, being compared to the case without defects and to the case of the uncovered design. The simulation corresponds to the design B dimer with $h = 200$ nm gold layer and $C = 1200$ nm. The resulting EF curves develop peaks that achieve higher values than the best obtained EFs in the uncovered design: see the black solid line.

Finally, we also simulate the presence of defects at the interface between the original cylinder and the covering layers. First, as shown in Fig. 8(a), the vacancies are present over a thickness $t$ centered at this interface. To create the pattern we start by randomly choosing vacancies on the surface with a density of one every $70 \times 70$ nm$^2$ and randomly grow the defect area and depth over several algorithm iterations. Colormaps corresponding to realizations of disorder for $t = 30$ nm and $t = 40$ nm are shown at the bottom of Fig. 8(a). We introduce these vacancies in the simulation of design B with covering length $C = 1200$ nm and layer height $h = 200$ nm. As seen in Fig. 8(b), the enhancement factor in presence of the vacancies gets reduced with respect to the case without imperfections. However, even for the large values of $t$ simulated, the EF peaks continue being superior to the largest values obtained in the uncovered design.

## 4 Conclusions

We have shown that the covered designs presented in Fig. 1 have the ability to reach large values of the enhancement factor surpassing the ones achieved in the uncovered design. This is true even in the presence of geometrical asymmetries in the dimer or vacancies at the interface between the nanorods and the covering layers. We find that the EF improvement arises as a result of a geometrical-induced rearrangement of the surface plasmon polaritons. This can be qualitatively understood by means of the location-resolved average charge densities defined in Section 2 having even number of nodes in the uncovered section with the top and bottom average densities becoming in phase at the peak conditions. We have discussed different approaches for the fabrication of these devices. Optimal choices of the covering length $C$ prove to be integer multiples of the SPP wavelength, here around 600 nm. For the uncovered length $U$ we have found that it is detrimental to the EF choosing values that would produce an EF peak in the uncovered dimer with length $U$ arms, indeed, the best choices are near the positions in which such uncovered dimer would develop an EF valley.

Our results show that, by geometrically engineering the gold covering added far from the hot-spot, covered designs can reach EF values larger than the highest ones obtained with uncovered designs. This finding, that could be extended to other dimer geometries, paves the way towards electrically contacting uncovered dimers while at the same time improving their performance in optical spectroscopy applications. In addition, due to their superior enhancement factors, non-contacted covering engineered designs can also expand the toolbox of high-performing SERS spectroscopy nanoantennas.

## Conflicts of interest

There are no conflicts to declare.

## Note added after first publication

This article replaces the version published on 3rd March 2021, which contained errors in eqn (1).

## Acknowledgements

We thank Eduardo Martinez for useful discussions. This work has been supported by the Consejo Nacional de Ciencia, Tecnología e Innovación Tecnológica del Perú (CONCYTEC), Contract No. 174-2018-FONDECYT-BM. A. A. R acknowledges additional support from CONICET (Argentina) and the Abdus Salam International Centre for Theoretical Physics (Italy). M. L. P. acknowledges the funding from FONCyT, PICT-2016-2531 (Argentina); and Universidad Nacional de Cuyo, SeCTyP, 344 C022 (Argentina). I. A. R. acknowledges support from 2018 M.Sc. scholarship CONCYTEC (Perú) contract No. 167.